\documentclass{PoS}

\title{The $t$-dependence of the pure DVCS cross section at COMPASS}

\ShortTitle{The $t$-dependence of the DVCS cross section}

\author{\speaker{P. Joerg} on behalf of the COMPASS collaboration\\ 
        University of Freiburg, Germany\\
        E-mail: \email{philipp.joerg@cern.ch}}

\abstract{The key reactions to study the Generalised Parton Distributions are Deeply Virtual Compton Scattering (DVCS) 
and Deeply Virtual Meson Production (DVMP). At COMPASS, these processes are investigated using a high intensity
muon beam with a momentum of 160\,GeV/c and a 2.5\,m-long liquid hydrogen target. In order to optimize the selection of exclusive reactions at these 
energies, the target is surrounded by a new barrel-shaped time-of-flight system to detect the recoiling particles.
COMPASS-II covers the up to now unexplored $x_{B}$ domain ranging from 0.01 to 0.15.
From the sum of cross sections measured with positive and negative beam polarities, the pure DVCS cross-section and its $t$-dependence have been extracted resulting
in a first model-independent determination of the transverse size of the partonic distribution of the nucleon $\sqrt{<r_{\perp}^2>}= (0.578 \ \pm \ 0.042\ _{- \ 0.018}^{+ \
0.006})\,\textsf{fm} $ at a mean $x_{B}$ value of 0.056.
}

\FullConference{XXIV International Workshop on Deep-Inelastic Scattering and Related Subjects\\
		11-15 April, 2016\\
		DESY Hamburg, Germany}

\newcommand{\PluLeft}{{\stackrel{\textstyle{+}}{\textstyle{\leftarrow}}}}
\newcommand{\MinRight}{{\stackrel{\textstyle{-}}{\textstyle{\rightarrow}}}}
\newcommand{\di}{{\rm d}}

\newcommand{\bitem}{\begin{itemize}}
\newcommand{\eitem}{\end{itemize}}
\newcommand{\be}{\begin{equation}}
\newcommand{\ee}{\end{equation}}
\newcommand{\bea}{\begin{eqnarray}}
\newcommand{\eea}{\end{eqnarray}}

\begin{document}

\section{Introduction}
Diffractive reactions are sensitive to the transverse size of the nucleon as a function of the Bjorken scaling variable $x_{B}$.
It is a natural consequence of Lorentz invariance (Ref. \cite{diehl}) that the transverse nucleon size, $r_{\perp}$, decreases as soon as $x_{B}$ starts to increase and a large
amount of the nucleon momentum is carried by less of its constituents.
However, the precise evolution of $r_{\perp}$ as a function of $x_{B}$, often referred as ``nucleon tomography'', is poorly known and
needs to be determined experimentally.
The dependence of the cross section of Deeply Virtual Compton Scattering,
 $\mu p \rightarrow \mu' p' \gamma$, as a function of the square of the 4-momentum transfer between initial and final state proton gives experimental
access to the transverse size of the nucleon (Ref. \cite{burkardt}).
Deeply Virtual Compton Scattering (DVCS), though experimentally challenging, is the most pure channel to study nucleon tomography, since in
contrast to Deeply Virtual Meson Production no final state interaction and no meson wave function has to be taken into account.
Extracting the slope $B$ of the exponential $t$-dependence of the t-differential DVCS cross section
$r_{\perp}(x_{B})$
can be defined as
\be
<r_{\perp}(x_{B})^2>=2 <B(x_{B})> \hbar^2.
\ee
The measurements of $B$ by HERA (Ref. \cite{zeus}-\cite{h1_2}) up to $x_{B}$ of approximately $10^{-3}$ do not show an evolution of $r_{\perp}(x_{B})$ in the kinematic range of the
sea quarks and gluons.
COMPASS is able to reach values of $x_{B}$ from 0.01 to 0.15 and thus gives new and vital input.
\newline
\newline
In this paper we present a measurement of the differential cross section
\be
\frac{\di^4 \sigma ^{\mu p \rightarrow \mu' p' \gamma }}{\di Q^2 \di \nu \di |t| \di \phi_{\gamma^* \gamma}}
\label{eq_x_sec}
\ee
for hard exclusive muoproduction of real photons off an unpolarised proton target at fixed incident muon energy.
Here $Q^2$ and $\nu$ denote the virtual photon virtuality 
and its energy respectively, $t$ the square of the total 4-momentum transfer between initial and final proton, 
and $\phi_{\gamma^* \gamma}$ the 
azimuthal angle between the lepton scattering plane and the photon production plane, according to the Trento convention.
For better readability the differential cross section of equation \ref{eq_x_sec} is abbreviated in the following by $\di \sigma$, while
the orientation of the muon charge and helicity are depicted by $\pm$ respectively $^{\rightarrow}_{\leftarrow}$.
The sum of the cross sections for $\mu^+$ or $\mu^-$ incident beams
\be
{\cal{S}}_{CS,U} \equiv \di\sigma^{\PluLeft} + \di\sigma^{\MinRight}
= 2 (\di\sigma^{BH} + \di\sigma^{DVCS}_{unpol} + e_\mu P_\mu \mathrm{Im} \; I)
\label{eq:ssc}
\ee
allows us to access the ``pure" DVCS contribution
after subtraction of the Bethe Heitler term. Furthermore recalling (Ref. \cite{belitsky})
\be
\di\sigma^{DVCS}_{unpol} = \frac{e^6}{y^2Q^2}
(c_0^{DVCS} + c_1^{DVCS} \cos \phi_{\gamma^* \gamma} + c_2^{DVCS} \cos 2 \phi_{\gamma^* \gamma})
\label{eq:DVCS_modulations}
\ee
and
\be
\mathrm{Im} \; I = \frac{e^6}{x_{B} y^3 t P_1(\phi)P_2(\phi)}
(s_1^{I} \sin \phi_{\gamma^* \gamma} + s_2^{I} \sin 2 \phi_{\gamma^* \gamma}))\textsf{,}
\label{eq:Int_modulations}
\ee
an integration of ${\cal{S}}_{CS,U}$ in $\phi_{\gamma^* \gamma}$ is performed for the cancellation of the interference term.
The lepton propagators are denoted by $P_1(\phi)$ and $P_2(\phi)$, while 
$c_i^{DVCS}$ and $s_i^{I}$ are combinations of Compton form factors and only
$c_0^{DVCS}$ and $s_1^{I}$ are the leading twist-2 contributions.
It should be underlined that the interference term has $\sin n \phi_{\gamma^* \gamma}$ modulations and thus an experimental acceptance symmetric around $\phi_{\gamma^* \gamma}=0$
supresses its contribution. Hence, only a contribution of the term $c_0^{DVCS}$ remains, which corresponds to the dominant transversely polarised virtual photon contribution of
the pure DVCS cross section.
The goal of this measurement is to 
measure the $|t|$-dependence of the pure DVCS cross section and to extract the slope parameter $B$.
The virtual photon-proton scattering cross section is derived from the measured muon-proton cross section by correcting for the virtual photon flux term:
\be
\frac{\di^4 \sigma ^{\mu p }}{\di Q^2 \di \nu \di |t|}
= \Gamma \Bigl(1+\epsilon \frac{\sigma_L}{\sigma_T}\Bigr) \frac{\di \sigma ^{\gamma^* p }}{ \di |t| },
\ee
where
\be
\Gamma = \frac{\alpha_{em}  (1- x_{Bj})}{2 \pi Q^2 y E_{\mu}} 
\left[ y^2\left(1 - \frac{2 m_{\mu}^2}{Q^2} \right)  + \frac{2}{1+\frac{Q^2}{\nu^2}} \left(1-y - \frac{Q^2}{4E_{\mu}^2} \right) \right]
\ee
for which the Hand convention is used.
Since only the transversely polarised virtual photon
contribution of the pure DVCS cross section is selected, the kinematic factor $\epsilon$ and the ratio of $\frac{\sigma_L}{\sigma_T}$ is not taken into account.

\section{Experimental setup}
COMPASS is a fixed target experiment at CERN. It can be operated with a tertiary $\mu^+$ or $\mu^-$ beam with an incident momentum of 100-200 GeV/c. The beam polarisation of $\pm$
80 \% for $\mu^{\mp}$ changes together with the beam charge. The COMPASS II detector is an open field two stage spectrometer. The beam is centered on a 2.5 m long
unpolarised liquid hydrogen target surrounded by two concentric rings of scintillating counters, which detect the recoil protons by means of the Time of Flight (ToF)
technique.
Directly downstream of the target a new electromagnetic calorimeter is placed.
The particles emitted in the forward direction are reconstructed by a two-stage magnetic spectrometer, equipped with 
a variety of tracking detectors, a muon filter for muon
identification and a hadronic as well as an electromagnetic calorimeter. A detailed description of the spectrometer can be found in Ref. \cite{proposal}.
\newline
The data shown in the following is extracted from a pilot run for the DVCS program, performed during four weeks in 2012. The major upgrades for this pilot run
are the proton recoil detector and the electromagnetic calorimeter located directly downstream of the target.
\section{Event selection}
The aim is to select exclusive single photon production, $\mu p \rightarrow \mu' p' \gamma$, 
while minimising the impact of events associated with background due to pile-up.
First, events containing 
a single outgoing muon and a single neutral cluster above the DVCS calorimeter thresholds are selected. The muon and photon are then combined with
all possible recoil proton candidates reconstructed inside the ToF detector surrounding the liquid hydrogen target. 
\begin{figure}[h!]
\begin{center}
\begin{tabular}{  c c }
\includegraphics[width=0.49\textwidth]{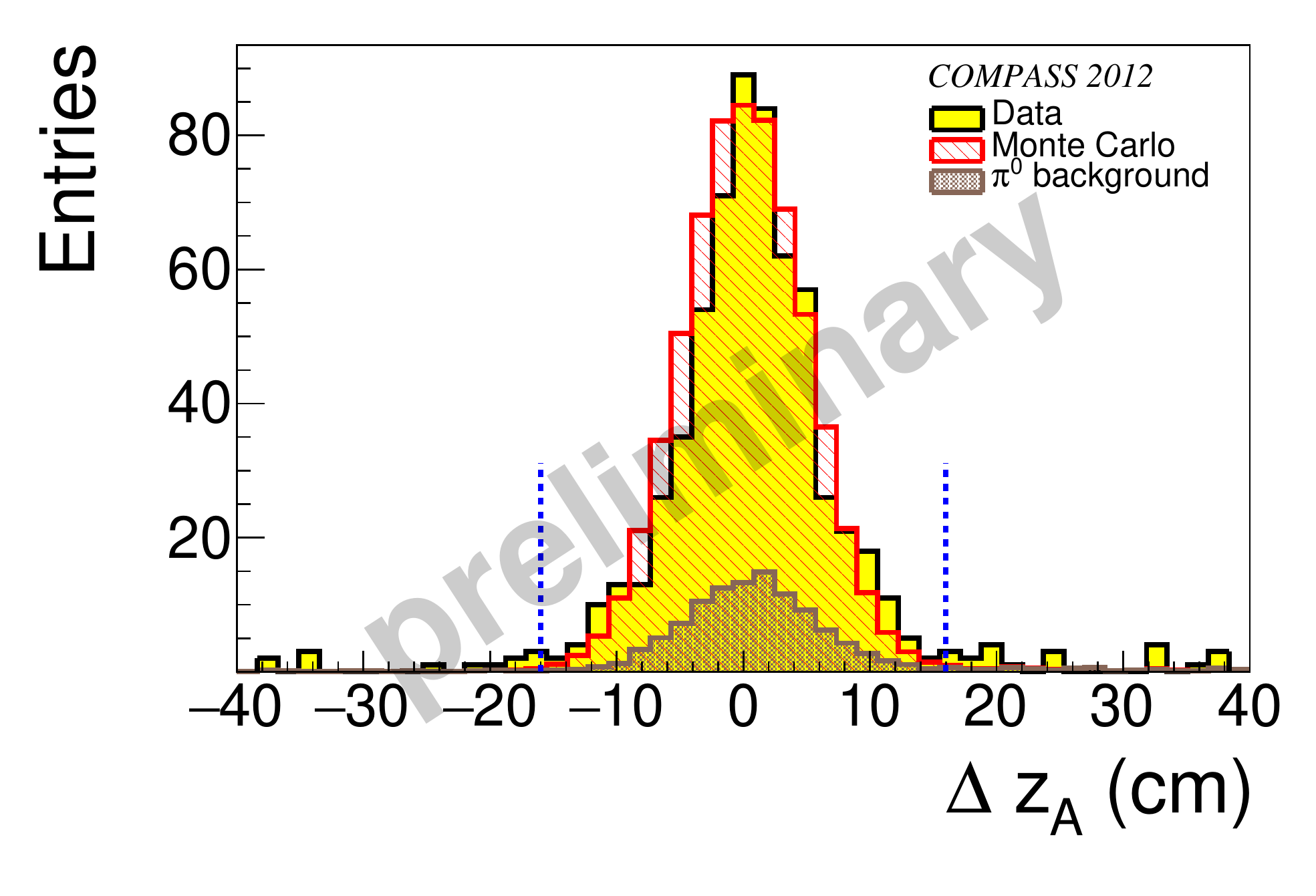}
&
\includegraphics[width=0.49\textwidth]{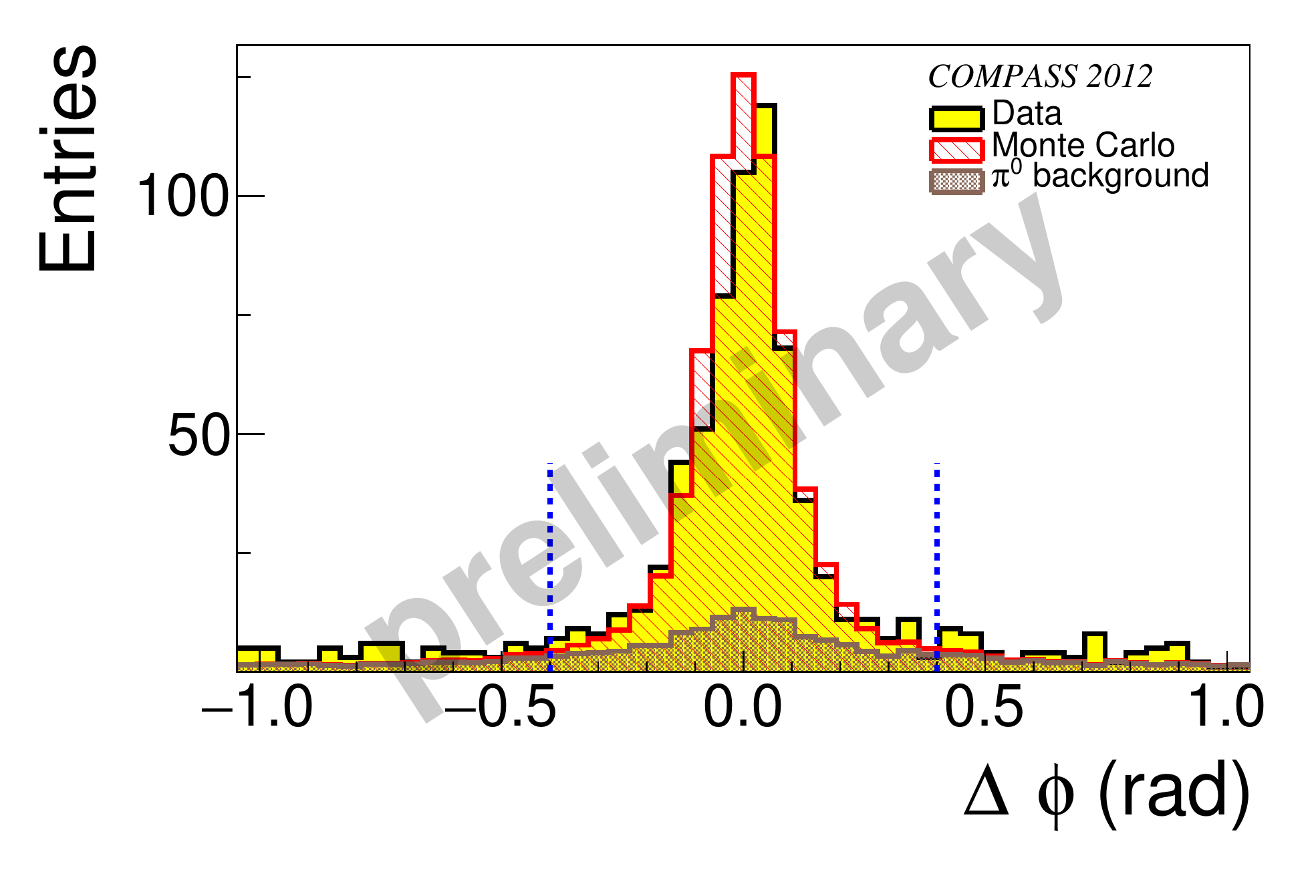}
\\
\includegraphics[width=0.49\textwidth]{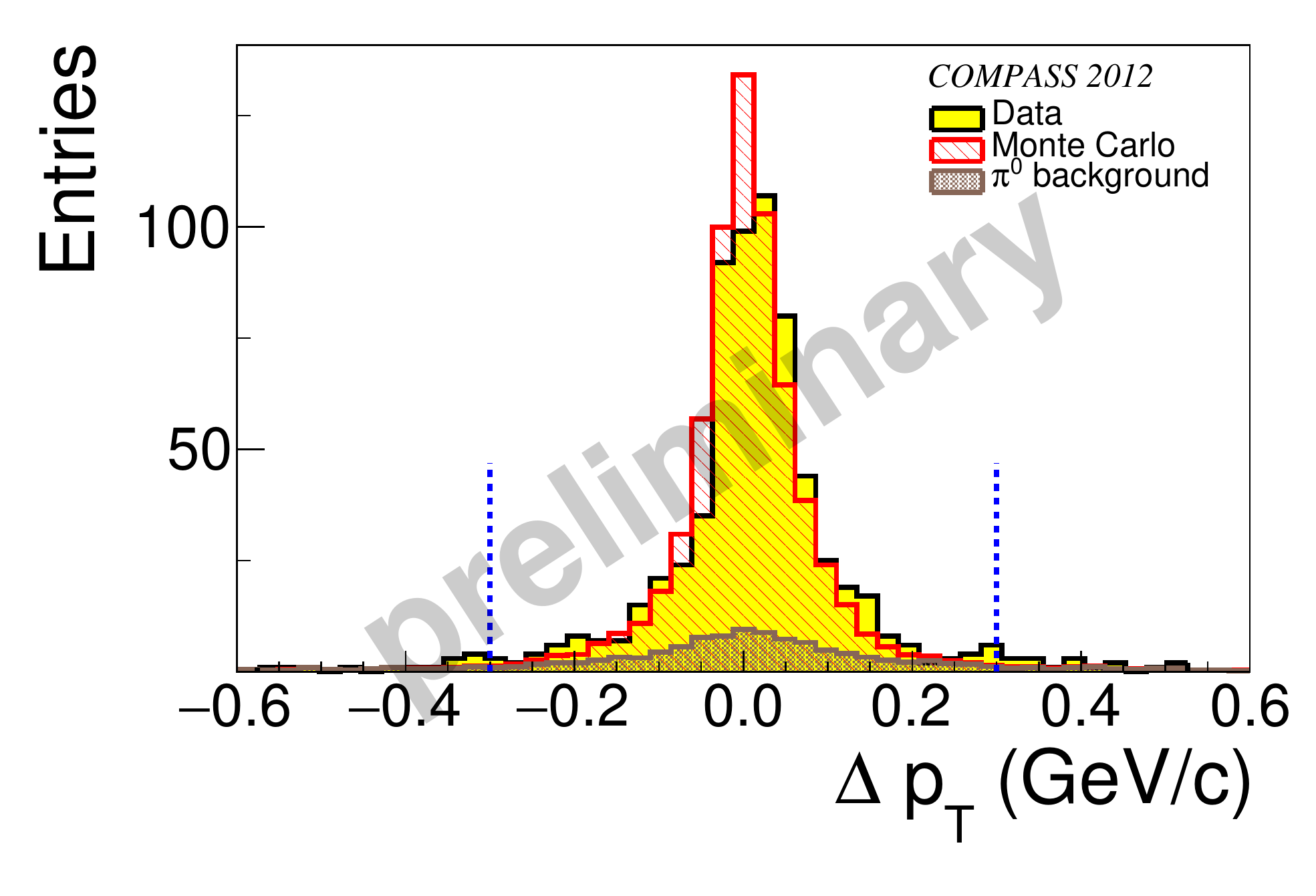}
&
\includegraphics[width=0.49\textwidth]{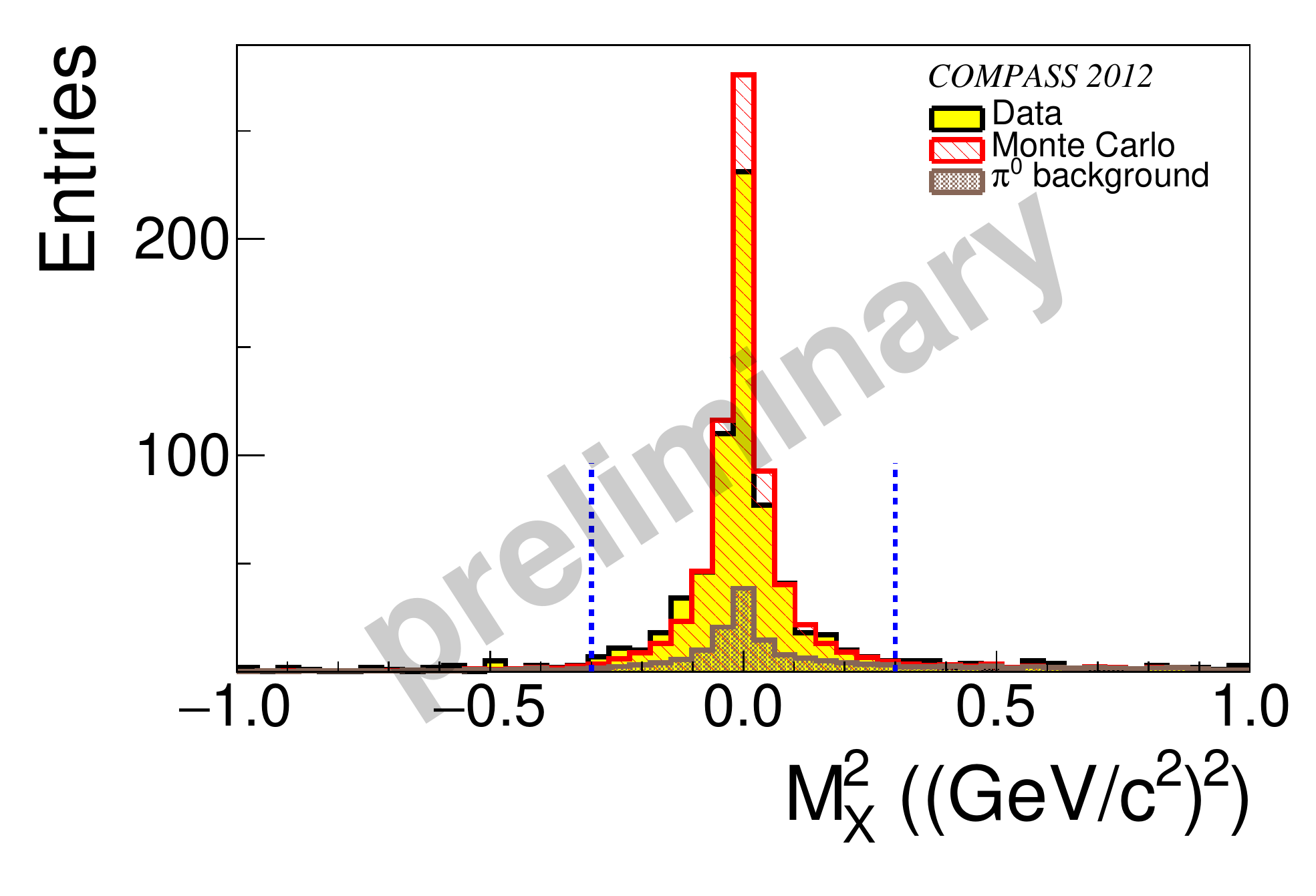}
\end{tabular}
\end{center}
\caption{\label{fig_exclusivity_cuts} 
Exclusivity variables: The whole Monte Carlo estimate is shown in red while in grey only the
$\pi^0$ contamination is shown. The variables $\Delta \varphi$, $\Delta p_{\bot}$ and $M_X^2$ are defined in equation \ref{eq_phi}-\ref{eq_m}
while $\Delta z_{A}$ encodes a reverse vertex pointing.
The blue dotted lines indicate the applied cuts.}
\end{figure}
The selection of exclusive events is further refined by applying cuts that take advantage of the over-constrained kinematics of the reaction: 
coplanarity
\be \Delta \varphi\ =\ \varphi_{meas}^{proton}\ -\ \varphi_{reco}^{proton}\textsf{,} \label{eq_phi} \ee
the transverse momentum balance
\be \Delta p_{\intercal}\ =\ p_{\intercal,meas}^{proton}\ -\ p_{\intercal,reco}^{proton}\textsf{,} \label{eq_p} \ee
as well as the total 4-momentum balance
\be M_X^2\ =\ (p_{\mu_{in}} + p_{p_{in}} - p_{\mu_{out}} - p_{p_{out}} - p_{\gamma})^2\ \label{eq_m}. \ee
The subscript ``meas" denotes measured quantities by the recoil detector
while the subscript ``reco" denotes that the same quantity has been calculated from the reconstructed incident and final
state muon together with the measured photon energy and the assumption of exclusivity.
\newline
In addition the longitudinal hit position $z_{A}$ inside the recoil detector, predicted from the interaction
vertex and the hit in the outer ring of the recoil detector, is compared to the measured hit position.
This reverse vertex pointing is encoded in the variable
$\Delta z_{A}$, which gives the difference of the two hit positions.

\section{Results}
\begin{figure}[h!]
\centering
\includegraphics[width=0.99\textwidth]{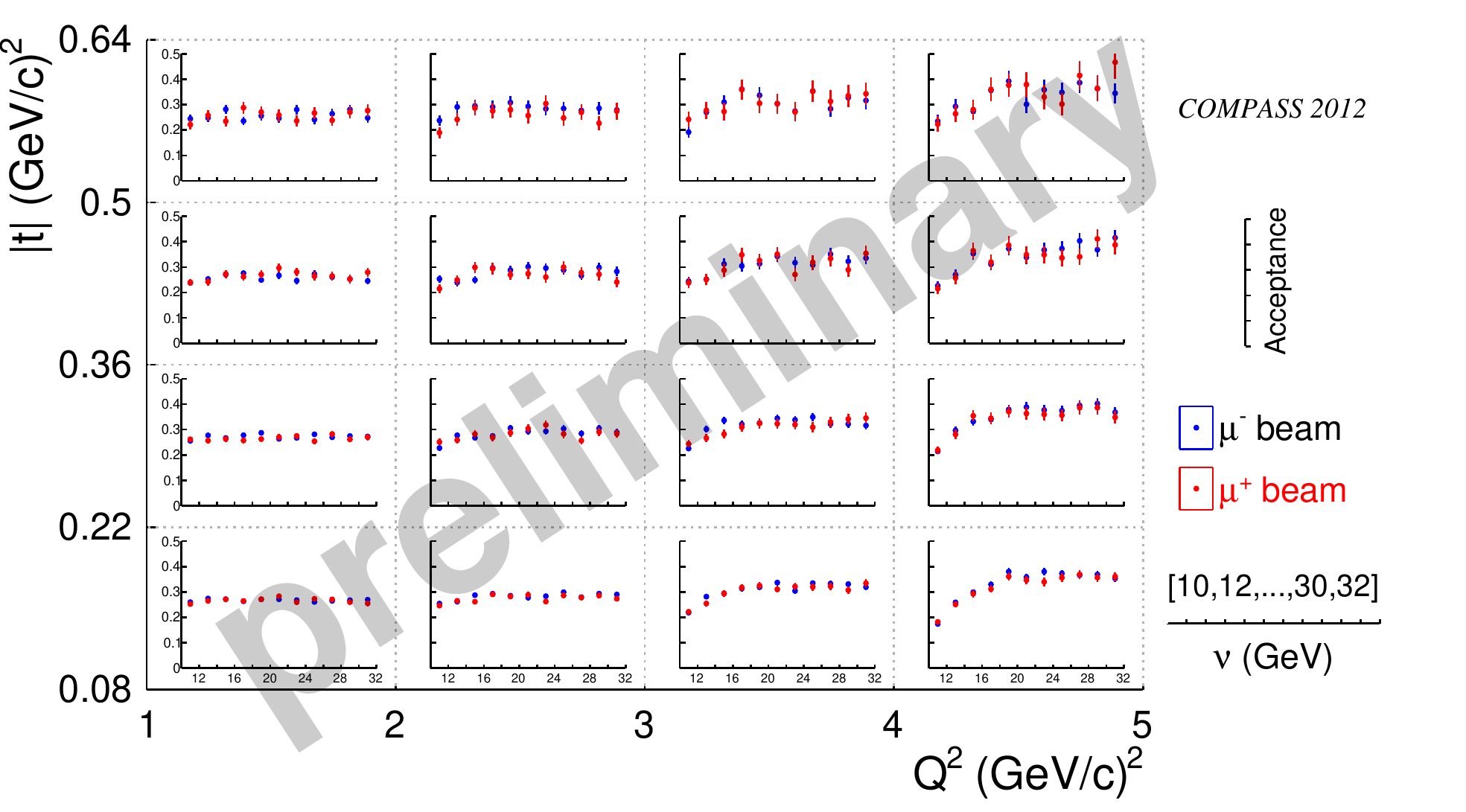}
\caption{Acceptance correction factors as a function of $t$, $Q^2$ and $\nu$.}
\label{acc}
\end{figure}
\begin{figure}[h!]
\centering
\includegraphics[width=.99\textwidth]{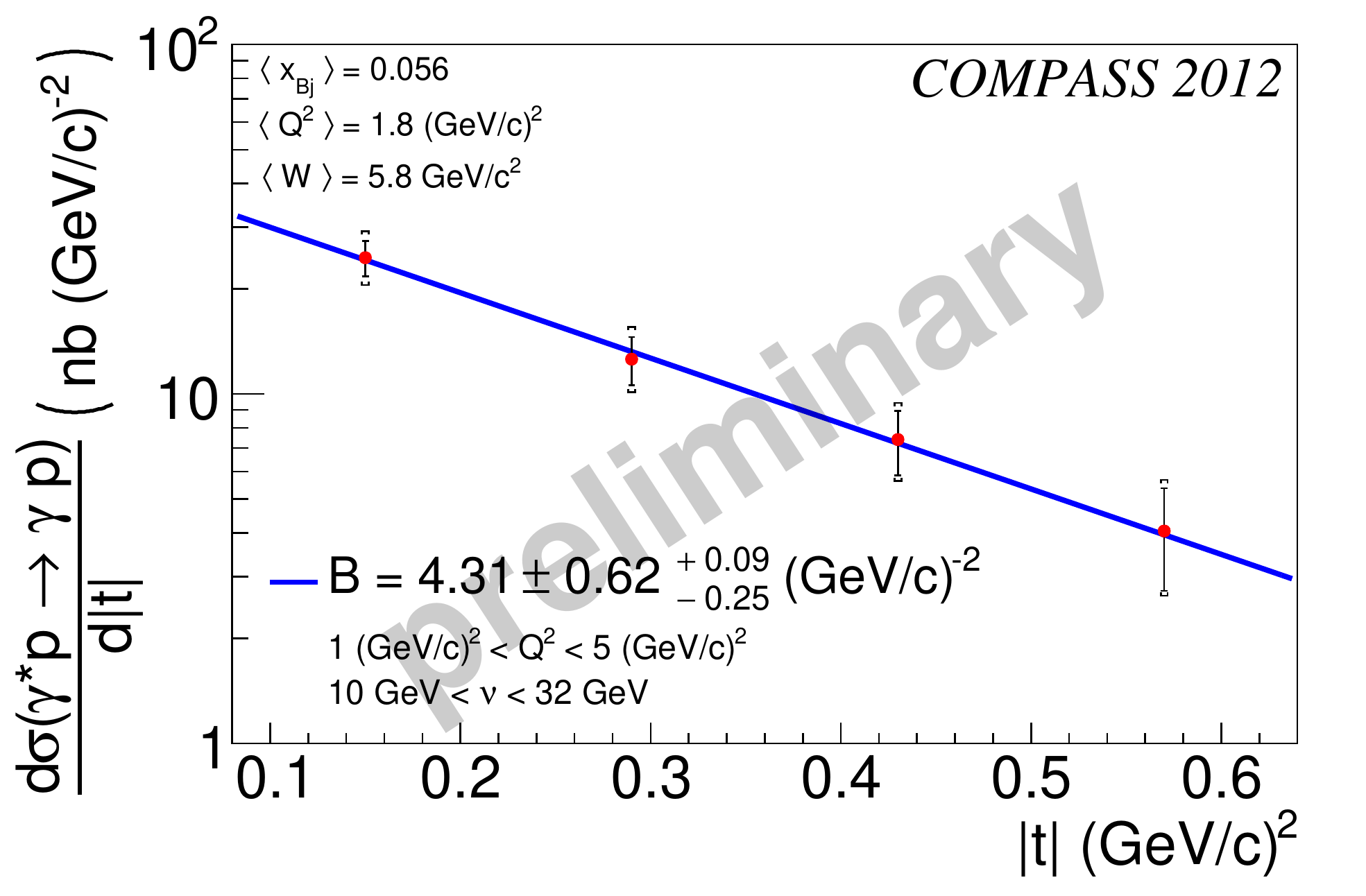}
\caption{The $t$-differential DVCS cross section as a function of $t$}
\label{x_sec}
\end{figure}

For the extraction of the cross section and the t-slope the data has been corrected for acceptance effects using binwise acceptance correction factors
according to figure \ref{acc}. The Bethe Heitler contribution has been subtracted from the data for each of these bins individually.
A major background source for single photon events is the production of pi0 mesons, where one of the two photons originating from the $\pi^0$ decay stays undetected.
The contamination arising from $\pi^0$ production was estimated and subtracted from the data inside each bin using LEPTO 6.5.1 together with an exclusive 
$\pi^0$ Monte-Carlo linked to a parametrisation from Goloskokov and Kroll (Ref. \cite{Kroll}).
A kinematically constrained fit, making full use of the exclusive measurement, has been
applied to obtain the most precise determination of the particle parameters at the interaction vertex.
The values of the differential cross section inside the four bins in $|t|$ have been fitted following an exponential law by using the binned maximum likelihood method to determine
the t-slope parameter. \newline The resulting parameter $B$ of the slope of the sum of the differential cross sections was determined to be
$B=(4.31 \ \pm \ 0.62\ _{- \ 0.25}^{+ \ 0.09})\,(\textsf{GeV}/\textsf{c})^{-2}$
inside the kinematic range
shown in figure \ref{x_sec},
which corresponds to mean kinematic values of 
$<x_{B}>=0.056$,  
$<Q^2>=1.8\, \textsf{(GeV/c)}^2$ 
and
$<W>=5.8\, \textsf{GeV}/\textsf{c}^2$.

The main systematic uncertainty on the cross section is linked to the absolute normalisation of the measurement, which results
in an upward uncertainty of up to 20 percent and a downward uncertainty of approximately 10 percent. Furthermore, the subtracted amount of $\pi^0$ background
can be translated into an additional upward uncertainty of up to 10 percent.
In case of the extracted parameter $B$ the systematic error originating from the absolute normalisation plays a minor role, while the
main systematic uncertainty of six percent originates from the absolute normalisation of the $\pi^0$ background and prefers a downward fluctuation.

\begin{figure}[h!]
\begin{center}
\begin{tabular}{  c c }
\hspace{-0.6cm}
\includegraphics[width=0.56\textwidth]{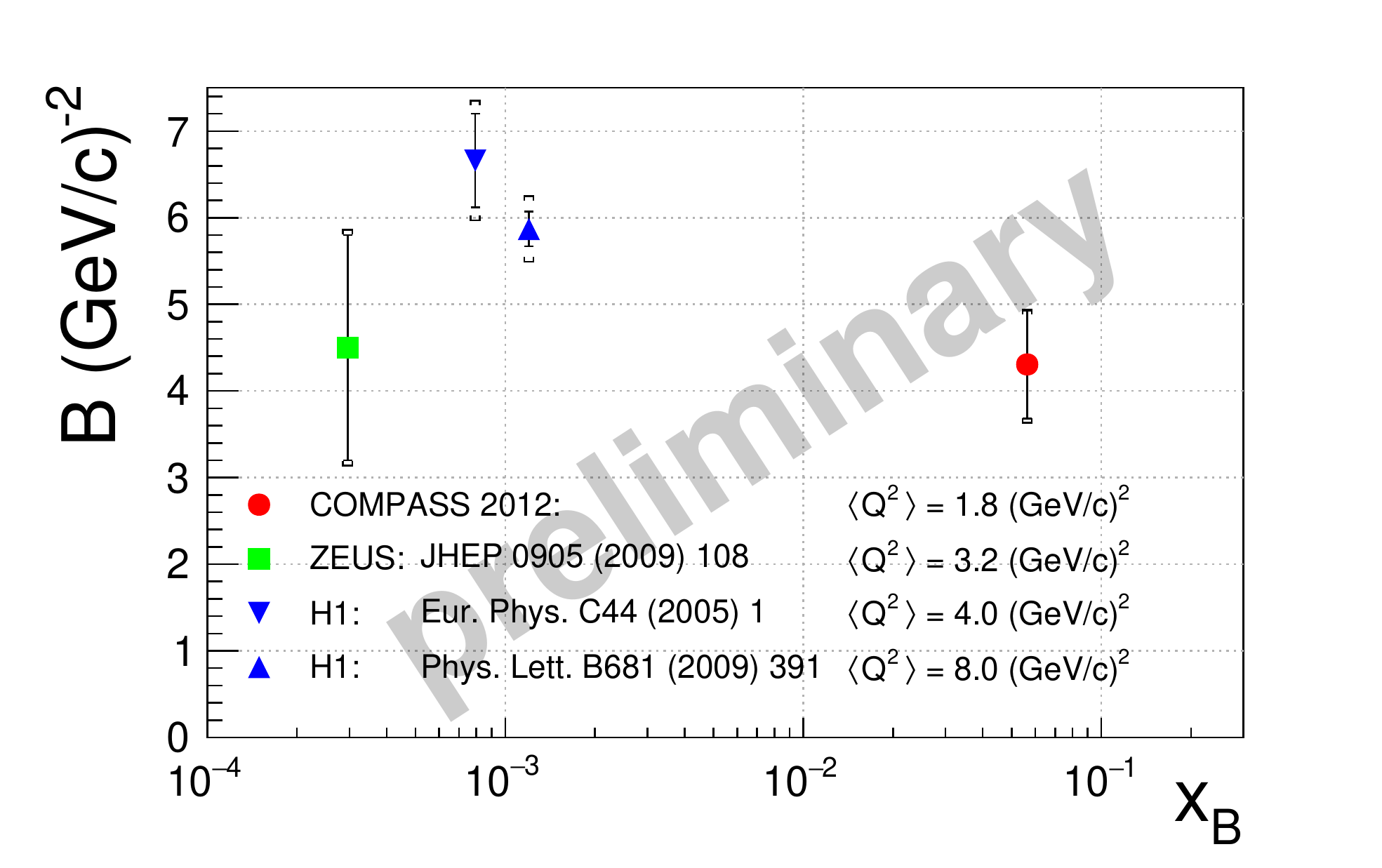}
&
\hspace{-1.3cm}
\includegraphics[width=0.56\textwidth]{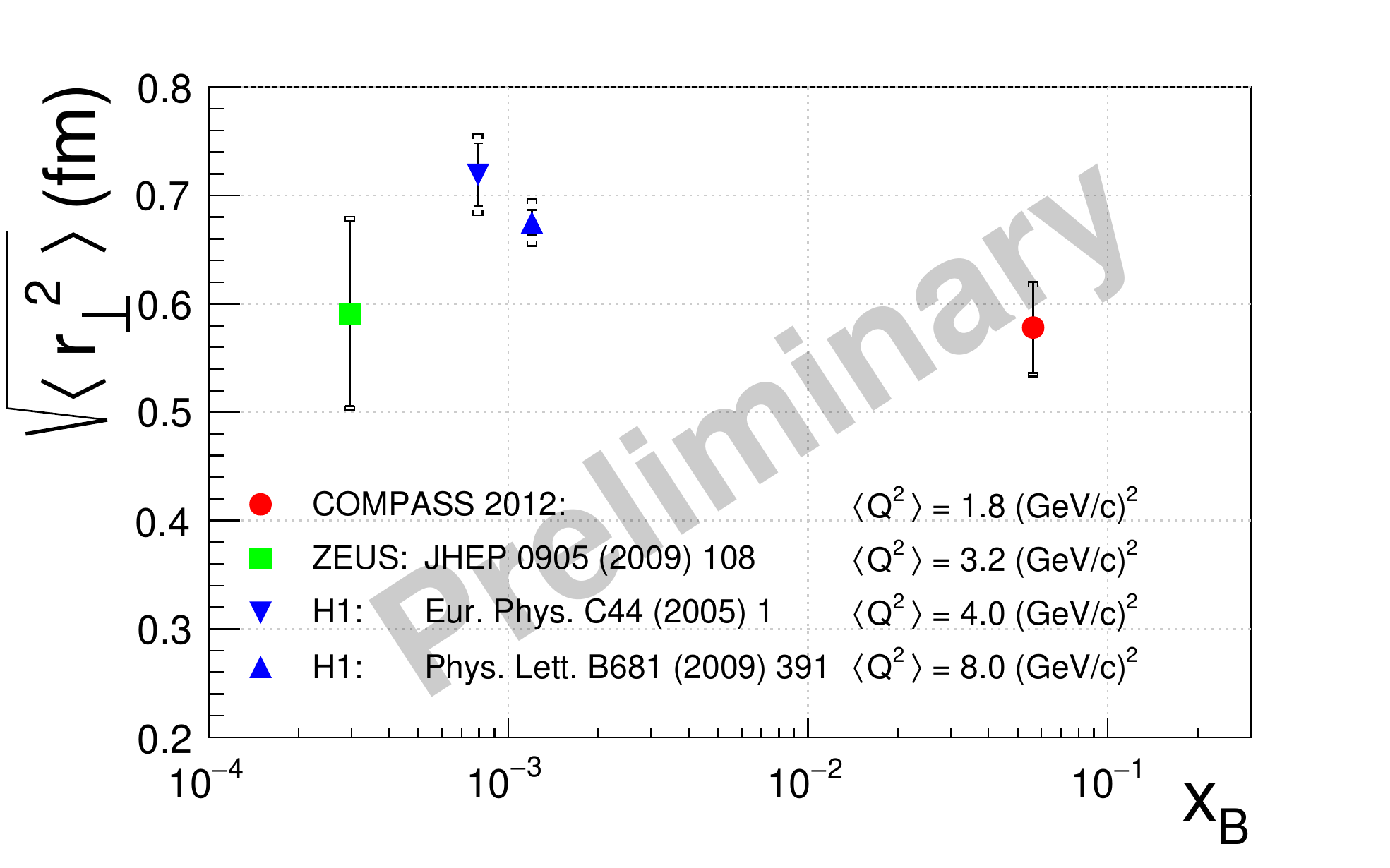}
\end{tabular}
\end{center}
\caption{\label{b_plot} 
COMPASS results for the t-slope parameter (\textbf{left}) and the extracted transverse size of the partonic distribution of the nucleon (\textbf{right}), compared with previous
HERA measurements.}
\end{figure}

Figure \ref{b_plot} shows the parameter $B$ and the resulting transverse size of the partonic distribution of a proton at $x_{B}=0.056$ extracted at COMPASS in comparison with results obtained by HERA.
This first model-independent determination of $\sqrt{ < r_{\perp}^2 > }$ may indicate a decrease of the transverse size of the proton as a function of the longitudinal momentum
fraction $x_{B}$. The statistics used in this analysis is approximately 7 percent of the data expected from the dedicated data takings in 2016 and 2017.
With the full statistics the granularity in $x_{B}$ will increase dramatically and the question whether a sizeable decrease of the transverse size of the proton can be observed in
this $x_{B}$ region will be answered.

\end{document}